\pdfoutput=1  % Instructs arXiv to run this with pdflatex rather than latex.
\documentclass[aps,twocolumn,groupedaddress,floatfix,preprintnumbers,nofootinbib,eqsecnum]{revtex4}
\usepackage{amssymb,amsmath,placeins,graphicx,multirow,xcolor}

\newcommand{\newc}{\newcommand}
\newc{\gsim}{\lower.7ex\hbox{$\;\stackrel{\textstyle>}{\sim}\;$}}
\newc{\lsim}{\lower.7ex\hbox{$\;\stackrel{\textstyle<}{\sim}\;$}}
\makeatletter
\newcommand{\biggg}{\bBigg@{3}}
\newcommand{\Biggg}{\bBigg@{4}}

\makeatother

\def\beq{\begin{equation}}
\def\eeq{\end{equation}}
\def\beqn{\begin{eqnarray}}
\def\eeqn{\end{eqnarray}}

\def\ie{{\it i.e.}\/}

\def\IR{\relax{\rm I\kern-.18em R}}
 \font\cmss=cmss10 \font\cmsss=cmss10 at 7pt
\def\IQ{\relax{\rm I\kern-.18em Q}}
\def\IZ{\relax\ifmmode\mathchoice
 {\hbox{\cmss Z\kern-.4em Z}}{\hbox{\cmss Z\kern-.4em Z}}
 {\lower.9pt\hbox{\cmsss Z\kern-.4em Z}}
 {\lower1.2pt\hbox{\cmsss Z\kern-.4em Z}}\else{\cmss Z\kern-.4em Z}\fi}

\def\OmegaDM{\Omega_{\mathrm{CDM}}}

\def\tnow{t_{\mathrm{now}}}

\def\MT2{M_{T2}}

\input epsf

\def\WMAP{Hinshaw:2012aka}
\def\DDMone{Dienes:2011ja}
\def\DDMtwo{Dienes:2011sa}
\def\DDMthree{Dienes:2012jb}
\def\DDMprocs{Dienes:2012iia,Dienes:2013qua,Dienes:2013rua}
\def\DDMHadrons{Dienes:2016vei,Dienes:2017ecw}
\def\DDMrandom{Dienes:2016kgc}
\def\anupam{Chialva:2012rq}
\def\DDMthermal{Dienes:2017zjq}
\def\DDMjeffone{Dienes:2015bka}
\def\DDMjefftwo{Dienes:2016zfr}
\def\DDMjeffproc{Dienes:2016mte}
\def\hiddenvalley{Strassler:2006im}
\def\hiddenvalleytwo{Essig:2009nc}
\def\DDMcollone{Dienes:2012yz}
\def\DDMcolltwo{Dienes:2014bka}
\def\DDMcollprocs{Dienes:2016udc}
\def\DDMdirectone{Dienes:2012cf}
\def\DDMAMS{Dienes:2013xff}
\def\DDMAMSprocs{Dienes:2014cca}
\def\DDMBoddyone{Boddy:2016fds}
\def\DDMBoddytwo{Boddy:2016hbp}
\def\DDMcomplementarityone{Dienes:2014via}
\def\DDMcomplementaritytwo{Dienes:2017ylr}
\def\MTtwo{Lester:1999tx}
\def\ATLASMonojet{Aaboud:2017phn}
\def\CMSMonojet{Sirunyan:2017hci}
\def\ATLASDijet{Aaboud:2017yvp}
\def\CMSDijet{Khachatryan:2015dcf}

\begin{document}

\title{Dynamical Dark Matter, MATHUSLA, and the Lifetime Frontier}
\author{
      David Curtin$^{1}$\footnote{E-mail address:  {\tt dcurtin@physics.utoronto.ca}},
      Keith R.\ Dienes$^{2,3}$\footnote{E-mail address:  {\tt dienes@email.arizona.edu}},
      Brooks Thomas$^{4}$\footnote{E-mail address:  {\tt thomasbd@lafayette.edu}} 
      }
\affiliation{
     $^1\,$Department of Physics, University of Toronto, Toronto, Ontario  M5S 1A7  Canada\\
     $^2\,$Department of Physics, University of Arizona, Tucson, AZ  85721  USA\\
     $^3\,$Department of Physics, University of Maryland, College Park, MD  20742  USA\\
     $^4\,$Department of Physics, Lafayette College, Easton, PA  18042  USA}
%  \date{\today}

\begin{abstract} 
\noindent
MATHUSLA is a proposed surface detector at CERN that would be able to observe the decays of non-hadronic electrically neutral long-lived particles (LLPs) with almost no background or trigger limitations.  This would allow MATHUSLA to probe sub-GeV to TeV masses and lifetimes up to $c\tau \sim 10^7~{\rm m}$. 
MATHUSLA can play an important role in probing dark-matter scenarios involving extended hidden sectors, where additional
dark states often manifest as LLPs.
A prime example of such a scenario is furnished by the Dynamical Dark Matter (DDM) framework, which intrinsically gives rise to large ensembles of dark states exhibiting a broad range of masses and lifetimes.
In this paper, we examine the extent
to which 
MATHUSLA can probe the DDM parameter space,
and we demonstrate
that MATHUSLA may be capable of providing direct confirmation of certain
unique aspects of the DDM framework which might be difficult to 
probe in other ways.
\end{abstract}

\maketitle

%========================================================================
%               MAIN TEXT BEGINS HERE
%========================================================================

%========================================================================

%\tableofcontents

\FloatBarrier
%%%%%%%%%%%%%%%%%%%%%%%%%%%%%%%%%%%%%%%%%%%%%%%%%%%%%%%%%%%%%%%%%%%%%%%%%%%%%%%%%%%%%%
\section{Introduction   \label{sec:Introduction}}
%%%%%%%%%%%%%%%%%%%%%%%%%%%%%%%%%%%%%%%%%%%%%%%%%%%%%%%%%%%%%%%%%%%%%%%%%%%%%%%%%%%%%%

Understanding the properties of dark matter is one of the most important outstanding mysteries of both particle physics and cosmology. (For recent reviews, see Refs.~\cite{Jungman:1995df,
Olive:2003iq, Hooper:2009zm, weinerlectures,Feng:2010gw, 
Lisanti:2016jxe, Battaglieri:2017aum}.)
Although much remains unknown about these properties,
the particles which constitute the dark matter must either be absolutely stable or ``hyperstable'', with
extremely long lifetimes $\tau \gsim 10^{26}$~s.
Methods of detecting dark matter can be classified as direct (involving the scattering of ambient dark-matter particles in a detector), indirect (involving astrophysical observation of the annihilation or decay of dark matter into
visible final states), or collider-based (involving the production of dark matter and its detection via missing-energy signatures).  
Only for indirect detection 
do the potentially finite lifetimes of the dark-matter particles themselves play a possible role. 

Since no human-scale experiment is capable of producing dark matter and directly observing the resulting decay, one might suspect that long-lived particle (LLP) searches can play no immediate role in the discovery or identification of dark matter. 
However, in many theories of dark matter, the stable or cosmologically long-lived particles which constitute the dark-matter energy density today are only part of an extended dark sector. 
Moreover, the other dark states residing within such extended sectors could couple to the Standard Model (SM) via highly suppressed interactions, and therefore have 
lifetimes which exceed collider timescales but which are nevertheless shorter than the age of the universe.
Such states are therefore potential LLPs.
The detection of such additional states is possible in ``hidden-valley'' scenarios~\cite{Strassler:2006im,Strassler:2006ri}, including theories of SIMP dark matter~\cite{Hochberg:2014dra}, ELDERs~\cite{Kuflik:2015isi}, co-decaying dark matter~\cite{Dror:2016rxc}, asymmetric dark matter~\cite{Zurek:2013wia}, and dark matter produced via cosmological 
freeze-in~\cite{Hall:2009bx}. Indeed, the prospects for detecting LLP states within these scenarios
are discussed in Ref.~\cite{Curtin:2018mvb}.

Dynamical Dark Matter (DDM)~\cite{\DDMone,\DDMtwo,\DDMthree,\DDMprocs} is an alternative 
framework for dark-matter physics in which the dark sector comprises 
an entire {\it ensemble}\/ of states exhibiting a broad range of lifetimes and cosmological abundances.
Those states within the ensemble whose lifetimes exceed the current age of the
universe together comprise the dark matter observed today.
Thus, within this framework, the fraction of the total energy density of the universe associated with dark matter 
is inherently dynamical, even in a matter-dominated epoch. 
Moreover, 
phenomenological considerations require that the lifetimes of the ensemble states
be carefully balanced against their cosmological abundances~\cite{\DDMone}, so that
states with larger abundances must have longer lifetimes but states with smaller abundances can
have shorter lifetimes.
This balancing therefore replaces the usual notion of dark-matter stability. 
Indeed, within this framework, the lifetimes and abundances of the states within the DDM ensemble are generally
connected 
through scaling relations which hold across the entire ensemble.
These scaling relations will be discussed further below.
Thus, the DDM framework is unique in connecting the finite lifetimes 
of possibly \emph{all} dark-sector states to the observed dark-matter relic abundance today. 
Theories of DDM are therefore natural targets of LLP searches at colliders such as the Large
Hadron Collider (LHC) or the future High-Luminosity Large Hadron Collider HL-LHC,
since they give rise to many different dark-sector LLPs exhibiting an entire spectrum of lifetimes,
potentially stretching from collider to cosmological timescales.

Since the DDM ensemble has to include states with lifetimes all the way up to the hyperstability bound $\tau \gsim 10^{26}$~s, it inherently motivates LLP searches in the long-lifetime regime. 
In general, collider acceptances scale as $\sim (L/\lambda)^n$ for 
a detector of size $L$, decay length $\lambda$, and number $n$ of observed decays needed for discovery.
Thus, within the long-lifetime regime with $\lambda \gg L$, only those LLP searches
involving a single LLP decay  are efficient.
However,
at the \mbox{(HL-)LHC} main detectors, searches for a single LLP are often limited by low trigger acceptances or complicated backgrounds, especially for LLPs that decay hadronically 
with less than a few hundred GeV,
or leptonically with less than $\sim 1-10$~GeV, of visible energy~\cite{Curtin:2018mvb}.

The recently proposed MATHUSLA detector~\cite{Chou:2016lxi} is designed to probe the long-lifetime regime 
by searching for displaced vertices on the surface above ATLAS or CMS.~  This detector will be able to discover LLPs with a similar geometric acceptance as the main detectors, but without background or trigger limitations. This can extend sensitivity by up to three orders of magnitude in cross section. The physics motivation for such a detector is very broad and general, as recently discussed in Ref.~\cite{Curtin:2018mvb}. 

In this paper, we shall demonstrate that MATHUSLA is uniquely well suited for probing DDM scenarios.
To do this, we shall evaluate the reach of the MATHUSLA detector within the parameter space of 
a particular DDM toy model.
Indeed, as we shall show, MATHUSLA can probe important regions of this parameter space that 
may be inaccessible to the main detectors.

This paper is organized as follows.
In Sect.~\ref{sec:MATHUSLA}, we describe the MATHUSLA detector and its overall capabilities.
Then, in Sect.~\ref{sec:DDM}, we provide 
a brief self-contained review of the salient features of the DDM
framework.
In Sect.~\ref{sec:colliders},
 we describe the potential signatures of DDM 
at colliders, and we sketch the ways in which MATHUSLA can play a unique and critical
role
in collider-based probes of DDM.~
In Sect.~\ref{sec:results}, we then examine the reach of the MATHUSLA detector within
the parameter space of our DDM toy model,
and find that there exist several compelling ``sweet spots'' within which MATHUSLA can be particularly
relevant for probing the corresponding dark sector. 
Finally, we conclude in Sect.~\ref{sec:Conclusions} with a discussion of possible
future directions for study.

\FloatBarrier
%%%%%%%%%%%%%%%%%%%%%%%%%%%%%%%%%%%%%%%%%%%%%%%%%%%%%%%%%%%%%%%%%%%%%%%%%%%%%%%%%%%%%%
\section{The MATHUSLA detector\label{sec:MATHUSLA}}
%%%%%%%%%%%%%%%%%%%%%%%%%%%%%%%%%%%%%%%%%%%%%%%%%%%%%%%%%%%%%%%%%%%%%%%%%%%%%%%%%%%%%%

In this section we briefly describe the details of the MATHUSLA detector that are relevant for our study. 
More details can be found in the original MATHUSLA paper~\cite{Chou:2016lxi}, in the MATHUSLA theory white paper~\cite{Curtin:2018mvb}, and in the MATHUSLA letter of intent~\cite{Alpigiani:2018fgd}.

MATHUSLA is motivated by the long-lifetime regime for LLPs, where main-detector searches are limited by complicated backgrounds and triggers that are optimized for prompt high-energy particle production. As a result, the inherently rare LLP signals in this regime can be swamped or overlooked, resulting in sensitivities for LLP decays which are much worse 
than implied by the geometric acceptance of the main detector.

MATHUSLA circumvents these limitations by operating on the surface above CERN, close to either ATLAS or CMS.~ We assume the benchmark geometry of Refs.~\cite{Chou:2016lxi, Curtin:2018mvb} (specifically, the so-called ``MATHUSLA200'' geometry 
discussed in Ref.~\cite{Alpigiani:2018fgd}), which assumes a 200$\,$m $\times$ 200$\,$m $\times$ 20$\,$m decay volume on the surface, displaced horizontally from the interaction point by 100~m, and instrumented only with a highly robust multi-layer tracking system above the decay volume. This allows for the reconstruction of displaced vertices for LLPs that decay inside the detector. Crucially, due to its large size, MATHUSLA can reconstruct displaced vertices in four dimensions, including consistent timing intersection of all upward-traveling charged tracks associated with the LLP decay. In principle, this allows all backgrounds, most importantly downward-traveling cosmic rays, to be rejected by the stringent displaced-vertex reconstruction criteria satisfied by the LLP signal.

In the long-lifetime regime, MATHUSLA has a similar geometric acceptance for LLP decays as the main detectors. However, since MATHUSLA is not hampered by trigger or background limitations, LLP searches at MATHUSLA can 
--- depending on the details of the LLP signal ---
potentially probe production cross sections that are several orders of magnitude smaller 
than can be probed through searches at the main detectors.
As discussed in detail in Ref.~\cite{Curtin:2018mvb}, the advantages of MATHUSLA are most pronounced for LLPs that decay hadronically with less than a few hundred GeV of visible final-state energy, or those that decay leptonically with masses below $\sim 1-10$~GeV. 
Since these include such generic scenarios as LLPs produced and decaying via the Higgs portal, MATHUSLA would clearly play a vital role in exploring the Lifetime Frontier. 
MATHUSLA can probe LLP cross sections $\lesssim$~fb for lifetimes in the 100$\,$m range. 
For $\sim$~pb LLP production cross sections, MATHUSLA can probe the BBN 
lifetime limit~\cite{Fradette:2017sdd} of $c \tau \lesssim 10^7$~m,
thereby scratching the ceiling 
of parameter space for most LLP theories. 
This capability is vital for maximizing the chances of discovering DDM.~

In our reach estimates, we assume perfect detection efficiency within the MATHUSLA decay volume and zero backgrounds. Since tracking efficiency is driven by cosmic-ray rejection and has to be excellent, this optimistic assumption is actually a very reasonable approximation of the true reconstruction efficiency for hadronically decaying LLPs~\cite{Alpigiani:2018fgd}. For leptonically decaying LLPs, the reconstruction efficiency will depend more on details of the final MATHUSLA design and tracking coverage of the decay volume, but these details would not qualitatively change our results.

\FloatBarrier
%%%%%%%%%%%%%%%%%%%%%%%%%%%%%%%%%%%%%%%%%%%%%%%%%%%%%%%%%%%%%%%%%%%%%%%%%%%%%%%%%%%%%%
\section{Dynamical Dark Matter\label{sec:DDM}}
%%%%%%%%%%%%%%%%%%%%%%%%%%%%%%%%%%%%%%%%%%%%%%%%%%%%%%%%%%%%%%%%%%%%%%%%%%%%%%%%%%%%%%

In this section we  outline the fundamental properties of the  DDM framework.  
We also explain why MATHUSLA may be able to provide evidence for this framework and 
potentially  place constraints on its overall properties.

We begin by recalling the traditional view of dark matter.
Within a traditional setup, the dark sector
is composed of one or several hyperstable dark-matter particle(s) $\chi$ which 
carry the entire dark-matter cosmological abundance
$\Omega_{\rm CDM} \approx 0.26$~\cite{\WMAP}.
This stability is critical for traditional dark matter. 
Indeed, any particle which decays too rapidly into SM 
states is likely to upset BBN and alter light-element abundances,
and also leave undesirable imprints in the CMB and
diffuse X-ray/gamma-ray backgrounds.
However, as a result of this stability,
the resulting dark sector is then essentially ``frozen'' in time, with
$\Omega_{\rm CDM}$ remaining constant in our late-time matter-dominated universe.
Moreover, as explained above, this stability also ensures that  once such a dark-matter particle  is produced in a collider, it escapes without any subsequent observable decay. 

Dynamical Dark Matter is different.
The DDM framework~\cite{\DDMone,\DDMtwo,\DDMthree,\DDMprocs} begins by assuming
that the dark sector consists of not merely one dark-matter particle, but {\it many}\/ such particles.
Indeed, the number $N$ of dark-matter particles can be relatively large, with $N$ reaching $10$, $100$, $1000$, or
even growing to infinity.
Thus, instead of having a single dark-matter particle $\chi$,
the dark sector contains an entire {\it ensemble}\/ of dark states $\chi_n$ ($n=0,...,N-1$).
Of course, no state individually needs to carry the full abundance $\Omega_{\rm CDM}$ so long as
the sum of their individual abundances  $\Omega_n$ matches $\Omega_{\rm CDM}$.
In particular, the individual dark components within the ensemble can carry a wide variety of
abundances $\Omega_n$, some relatively large but others relatively small.
This is a critical observation, because
{\it a given dark-matter component $\chi_n$ need not be stable if its
abundance $\Omega_n$ at the time of its decay into SM states is sufficiently small}\/.
Indeed, a sufficiently small abundance assures that all of the disruptive effects
of the decay of $\chi_n$ into SM states will be minimal, and that all
constraints from BBN, CMB, {\it etc.}\  will continue to be satisfied.

We are thus naturally led to an alternative concept~\cite{\DDMone}:   {\it a balancing of 
decay widths $\Gamma_n$ against cosmological 
abundances $\Omega_n$}\/, where $\Gamma_n$ are the widths for decays into SM states.
Dark-matter states with larger abundances must have smaller decay widths
and survive until (and potentially beyond) the present time,
whereas states with smaller abundances can have larger decay widths and decay at earlier times.
As long as decay widths are balanced against abundances in this way across our entire
dark-sector ensemble, all phenomenological constraints can potentially be satisfied.
Thus, dark-matter hyperstability is no longer required.

This, then, is DDM:  
an alternative framework for dark-matter 
physics in which the notion of dark-matter stability is
replaced by a balancing of lifetimes against cosmological abundances
across an ensemble of individual dark-matter 
components $\chi_n$ with
different masses $m_n$, lifetimes $\tau_n\equiv \Gamma_n^{-1}$, and cosmological abundances $\Omega_n$.
In some sense, this is the most general dark sector that can be contemplated,
reducing to the standard picture of a single stable particle as
$N\to 1$.
However, as $N$ is increased from $1$, 
we now see that the notion of dark-matter
stability generalizes into something far richer:
a balancing of lifetimes against abundances.  
In other words, 
the dark sector becomes truly {\it dynamical}\/,
with the different components
of the DDM ensemble decaying before,
during, and after the present epoch.
Indeed, some portions of the DDM ensemble
have already decayed prior to the present epoch,
and are thus no longer part of the dark sector.
However, other portions of the DDM ensemble have yet to decay.
It is these ensemble constituents whose abundances $\Omega_n$ together comprise
the specific dark-matter abundance $\Omega_{\rm CDM}\approx 0.26$ observed today.

Since the original DDM proposal~\cite{\DDMone,\DDMtwo,\DDMthree},
there have been many explicit realizations of such DDM ensembles --- \ie, many different theoretical
scenarios for physics beyond the SM which naturally give rise to a large collection of dark states in which
the widths for decays into SM states are naturally inversely balanced against cosmological abundances.
These include theories involving large extra spacetime
dimensions~\cite{\DDMone,\DDMtwo,\DDMthree}, theories involving strongly-coupled
hidden sectors~\cite{\DDMHadrons}, theories involving large spontaneously-broken
symmetry groups~\cite{\DDMrandom},
and even string theories~\cite{\DDMHadrons,\anupam}.
Indeed the dark states within these different realizations
can accrue suitable cosmological abundances in a variety of
ways, including not only through non-thermal generation 
mechanisms such as misalignment production~\cite{\DDMone,\DDMtwo,\DDMthree}
but also through thermal mechanisms such as freeze-out~\cite{\DDMthermal}. 
Mass-generating phase transitions
in the early universe can also endow collections of such states with non-trivial
cosmological abundances~\cite{\DDMjeffone,\DDMjefftwo,\DDMjeffproc}.

In these and other realistic DDM scenarios, the masses, lifetimes,
and abundances of these individual particles are not arbitrary.
Rather, these quantities follow directly from the underlying physics model
and generally take the form of {\it scaling relations}\/
(either exact or approximate)
which dictate how these quantities scale relative to one another across 
the ensemble as a whole.  
Through these scaling relations, the properties of the
ensemble constituents --- and thus the properties of the ensemble itself ---
are completely specified through only a small number of free parameters.
Thus, even though the number of particles which
contribute to the total dark-matter abundance is typically quite large,
specific top-down realizations of the DDM framework are very predictive.

The most fundamental of these scaling relations governs the spectrum
of masses for the DDM constituent particles $\chi_n$. 
In general, we assume a constituent mass spectrum of the form
\beq
         m_n ~=~ m_0 + (\Delta m) \, n^\delta
\label{masses}
\eeq
where $\lbrace m_0, \Delta m, \delta\rbrace$ are arbitrary parameters
and where $\Delta m, \delta  > 0$ 
(so that $n$ labels the DDM constituents in order of increasing mass).
Indeed, most concrete realizations of DDM ensembles have mass spectra which take
this general form, either exactly or approximately. 
For example, if --- as in Refs.~\cite{\DDMone,\DDMtwo} --- 
the ensemble constituents are the Kaluza-Klein (KK) excitations of a scalar
field compactified on a circle of radius $R$ (or a $\IZ_2$ orbifold thereof), 
we have either $\lbrace m_0,\Delta m,\delta\rbrace = \lbrace m, 1/R, 1\rbrace$
or $\lbrace m_0,\Delta m,\delta\rbrace =\lbrace m, 1/(2 m R^2), 2\rbrace$,
depending on whether $m R \ll1$ or $mR\gg 1$, 
respectively, 
where $m$ is the four-dimensional scalar mass.
In general, for arbitrary $m R$, we find that the latter behavior
holds for $n\ll mR$ and the former for $n\gg mR$.
Likewise, if the ensemble constituents consist of the bound states of a strongly-coupled
gauge theory, as in Refs.~\cite{\DDMHadrons}, we have $\delta = 1/2$, where $\Delta m$ and $m_0$ are related to
the Regge slope
and Regge intercept of the strongly-coupled theory, respectively.
Thus $\delta=1/2$, $\delta=1$, and $\delta=2$ may be considered as particularly compelling ``benchmark'' values.

Given a mass spectrum of this general form, we then typically take a scaling
relation for the decay widths $\Gamma_n$ of the form
\beq
            \Gamma_n ~=~ \Gamma_0 \left( {m_n\over m_0}\right)^y
\label{decaywidths}
\eeq
where $\Gamma_0$ is the decay width of the lightest DDM state and where $y$ is an additional free parameter.   
Note that $\Gamma_n$ is assumed to be the decay width of the $n$th ensemble constituent $\chi_n$
into SM states, and in such analyses one typically disregards the possibility of intra-ensemble decays
(or assumes that the branching ratios for such decays are relatively small).
The corresponding $\chi_n$ lifetimes are then given by $\tau_n \equiv \Gamma_n^{-1}$.
In general, the scaling exponent $y$ can be arbitrary.
For example, if we assume that the dominant decay mode of $\chi_n$ is to a final state consisting of SM particles whose masses are all significantly less than $m_n$, and if this decay occurs through a dimension-$d$ contact operator of the form ${\cal O}_n \sim c_n \chi_n {\cal O}_{\rm SM}/\Lambda^{d-4}$ where $\Lambda$ is an appropriate mass scale and where ${\cal O}_{\rm SM}$ is an operator built from SM fields, we have
\beq
                   y ~=~ 2d - 7~.
\label{yd}
\eeq
Likewise, if we assume that the $\chi_n$ are dark hadrons experiencing 
``hidden-valley''~\cite{\hiddenvalley,\hiddenvalleytwo} 
decays mediated by a dark photon, we would have $y=5$. 
In general one finds $y>0$, but this is not a strict requirement.
Indeed, since the fundamental couplings that underlie such decays can often themselves depend on $n$,
the scaling exponents $y$ can often grow quite large.

There are also additional important scaling relations which are often utilized in the DDM literature.
For example, one important quantity for many purposes is the spectrum of cosmological abundances   
$\Omega_n$ associated with each DDM constituent.  These are likewise assumed to satisfy an approximate
scaling relation of the form
\beq
                   \Omega_n ~=~ \Omega_0 \left( {m_n\over m_0}\right)^\gamma~.
\eeq
The precise value of the scaling exponent $\gamma$ generally depends on the particular 
dark-matter production mechanism
assumed.
One typically finds that $\gamma<0$ for misalignment production~\cite{\DDMone,\DDMtwo}, while
$\gamma$ can generally be of either sign for thermal freeze-out~\cite{\DDMthermal}.

In a similar vein, for many investigations 
it proves important to focus on the coupling coefficients $c_{m,n,...p}$ of Lagrangian operators
which involve multiple ensemble constituents $\lbrace \chi_{m},\chi_{n},...,\chi_p\rbrace$
together with some set of particles outside the ensemble.
Such couplings can ultimately be relevant for 
dark-matter production, scattering, annihilation, and decay.
In the analysis below, we shall mostly be interested in couplings that involve two dark-matter constituents $\chi_m$ and $\chi_n$ (or their antiparticles),
and we shall further restrict our attention to the ``diagonal'' case in which $m=n$. 
We shall then assume a scaling relation 
for such couplings of the form
\beq
                   c_{nn} ~=~ c_0 \left( {m_n\over m_0}\right)^\xi
\label{couplings}
\eeq
where $c_0$ is an overall normalization and where $\xi$ is a corresponding scaling exponent.
Assuming a scaling relation of this general form allows us to study a wide variety of underlying theoretical mechanisms
that might generate such couplings.   
For example, 
$\xi=0$ corresponds to democratic decay into different final states that are much lighter than the parent
particle, while $\xi=1$ corresponds to a Yukawa-like coupling.
Of course, once a particular scaling relation for the coupling is specified, the scaling behaviors of the 
corresponding production, scattering, or annihilation cross sections 
are also determined.
Since these cross sections also depend on kinematic factors, their behavior across the ensemble can
deviate significantly from the kinds of simple power-law relations we have assumed for the couplings.
For example, the results in Ref.~\cite{\DDMthermal} --- although derived for an entirely different purpose --- 
can be interpreted as illustrating the tremendous range of possible scaling behaviors
that can be exhibited by an
annihilation cross section 
even when the underlying annihilation couplings $c_{nn}$
are held fixed, with $\xi=0$, across the entire DDM ensemble.

In general, the phenomenological viability of the DDM framework rests upon certain  
relations between these different scaling exponents.  Two of the most important which underpin
the entire DDM framework are the relations~\cite{\DDMone,\DDMthermal}
\beq
            {\gamma y } ~<0~
\eeq
and 
\beq
           -1 ~\lsim~ {1\over y} \left( \gamma + {1\over \delta}\right) ~< ~ 0~.
\eeq
The first of these relations ensures a proper balancing of lifetimes against abundances
across the entire DDM ensemble, as discussed above. 
By contrast, the second relation
ensures a suitable equation of state for the  
collective DDM ensemble, 
with an effective equation-of-state parameter $w_{\rm eff} \approx 0$ which does not change appreciably 
over a significant portion of the recent cosmological past~\cite{\DDMthermal}.
Moreover, in cases of DDM ensembles with infinite numbers of closely-spaced constituents, 
this relation also ensures that the total energy density 
carried by the ensemble is finite.

%======================================
\section{Dynamical Dark Matter at colliders, and the role of MATHUSLA\label{sec:colliders}}

Scenarios within the DDM framework can give rise to distinctive signatures at
colliders~\cite{\DDMcollone,\DDMcolltwo,\DDMcollprocs}, at direct-detection
experiments~\cite{\DDMdirectone}, and at indirect-detection experiments~\cite{\DDMAMS,\DDMAMSprocs,\DDMBoddyone,\DDMBoddytwo}.
Such scenarios also give rise to enhanced complementarities~\cite{\DDMcomplementarityone,\DDMcomplementaritytwo} 
between different types of experimental probes.

In this section, we discuss the DDM signatures that are possible at the HL-LHC, 
assuming the existence of a production channel for the DDM ensemble.
While a variety of prompt signals are possible, 
we focus on the features most relevant to the long-lifetime regime:  missing energy and LLP signatures. 
In doing so, we highlight the important role of the LLP search program in general 
and hence the important role MATHUSLA can play in collider-based searches for DDM.

%=====================================
\subsection{DDM at colliders:  MET signatures}

As a result of the variety of decay lifetimes exhibited by the 
different components $\chi_n$ of a generic DDM ensemble,
these components may manifest
themselves in qualitatively different ways at colliders --- even in situations
in which the $\chi_n$ all have similar quantum numbers and are therefore
produced via similar processes.  

If the shortest lifetime in the DDM ensemble is $c \tau \gtrsim 10~\mathrm{m}$,
a significant fraction or all of the DDM states produced at the LHC escape the detector and can only be reconstructed as missing energy (MET).~ 
In that case, the scenarios with the best detection  prospects at the main detectors involve DDM states that are produced with sizable cross sections and in association with prompt visible final states to ensure a large MET signal. 

For example, techniques have been developed in Refs.~\cite{\DDMcollone,\DDMcolltwo,\DDMcollprocs} 
for discerning the existence
of an entire DDM ensemble of dark-matter components at hadron colliders such as the LHC
or future high-luminosity \mbox{(HL-)LHC},
and for distinguishing such DDM ensembles 
from more traditional dark-matter candidates.
Such studies focused on DDM scenarios in which the
ensemble constituents are produced via the decays of additional, heavier ``parent'' particles which are charged under
$SU(3)_c$ color and can therefore be pair-produced copiously at hadron colliders.  In cases in which each
parent particle decays to a single ensemble constituent and a pair of hadronic jets, it was shown 
in Ref.~\cite{\DDMcollone}
that the invariant-mass
distribution of these two jets could provide a distinctive signal of DDM.~
Indeed, this possibility extends 
throughout large portions of the underlying DDM parameter space of scaling exponents~\cite{\DDMcollone}.
Moreover, decay topologies of this sort arise generically
in cases in which
both the parent particle and the constituents of
the DDM ensemble are charged under an approximate symmetry. 
Similarly, in cases in
which the parent particle decays primarily into a single ensemble constituent and a single jet, it was shown 
in Ref.~\cite{\DDMcolltwo}
that the $M_{T2}$ distribution~\cite{\MTtwo} can likewise provide such a signal.

The general lesson from such studies is that
distinguishing between minimal and DDM-like non-minimal dark sectors at colliders typically
involves more than merely identifying an excess in the total number of signal events
over background.  In particular, it typically requires a detailed analysis of the
shapes of relevant kinematic distributions.  

It is also important to note that in comparison with simple, ``bump-hunting'' searches,
distribution-based searches of this sort involve a number of additional subtleties~\cite{\DDMcolltwo}.
For example, cuts imposed on the data
for purposes of background reduction can distort event-shape distributions
whenever non-trivial correlations exist between the corresponding collider variables.
The results of an examination of the effects of 
these distortions on the event-shape distributions of the kinematic variables most sensitive
to the structure of the dark sector were reported in Ref.~\cite{\DDMcolltwo}.
Indeed, it was shown that appropriately chosen cuts on certain variables
can actually enhance the distinctiveness of these distributions, while
cuts imposed on other variables 
tend to obscure signals of
non-minimality in the dark sector.
More details behind all of these results can be found in
Refs.~\cite{\DDMcollone, \DDMcolltwo}.

Such scenarios for DDM are somewhat optimistic in terms of their collider-based detection prospects
and illustrate how much information about the ensemble 
can be extracted if the production cross section and missing energy per event are large. 
However, even scenarios which are much less favorable in terms of their collider-detection prospects
can be effectively probed using monojet~\cite{\ATLASMonojet,\CMSMonojet} and other related searches that exploit the inevitable MET signal from initial-state radiation associated with any potential DDM 
or other invisible particle-production process. 
Recent projections for the reach of monojet searches at the HL-LHC can be found in Ref.~\cite{Curtin:2018mvb}. 
For typical benchmark scenarios where an SM-singlet mediator has couplings $\lesssim 1$ to SM fermions and a 
single dark-matter species, mediator masses above a TeV can be discovered at the HL-LHC.~ 
In this case, less information about the ensemble may be available, since the MET spectrum is 
dictated by the properties of the mediator and not the invisible particle(s). 
This nevertheless demonstrates the possible reach of the HL-LHC in probing DDM and other dark-matter scenarios. 
Of course, if the DDM ensemble includes states that are sufficiently short-lived to yield even just a few decays inside MATHUSLA or the main detectors, then the observation of those decays could provide further evidence for DDM, especially if a multitude of masses and lifetimes is observed and if those measurements are correlated with the results of MET searches to further elucidate the properties of the DDM spectrum.

%================
\subsection{DDM at colliders: LLP signatures}

The direct observation of several different LLPs obeying discernible scaling relations constitutes an obvious smoking gun for the DDM framework. If all of the observable decay lengths are similar to or smaller than the LHC main-detector size, the analysis strategy involves correlating the discoveries made in several different main-detector searches to uncover the properties of the ensemble.  

But what if we do not have such an embarrassment of riches?  
Indeed, this would be the case in the long-lifetime regime. 
Understanding how to probe DDM in this regime is important for purely pragmatic reasons, 
but also has a particular theoretical motivation. 
The DDM ensemble must contain states with lifetimes up to the hyperstability bound. Hence, if 
the DDM ensemble contains states with {\it any}\/ collider-observable lifetime, then it is 
likely to also contain states all the way up to {\it the upper bound}\/ of 
collider-observable lifetimes.  Therefore, while a search program specifically targeting 
the long-lifetime regime is important for many different scenarios for physics beyond the 
SM, it is particularly well-motivated for DDM.~ 

That said, searching for LLPs is challenging at the main LHC detectors, as detection 
prospects at these detectors are significantly limited by triggering requirements and 
complicated backgrounds.  By contrast, as discussed in Sect.~\ref{sec:MATHUSLA}, 
MATHUSLA does not suffer from these limitations and is therefore capable of providing 
a far greater discovery reach in LLP searches --- often by orders of magnitude --- than 
the main LHC detectors, despite having a comparable geometric acceptance.  For this
reason, MATHUSLA affords potentially the only discovery opportunity for many 
new-physics scenarios involving LLPs --- especially those in which the LLPs are 
relatively light and decay hadronically.  In the context of the DDM framework,
these considerations are doubly important, since probing the underlying structure of the 
DDM ensemble requires the observation of several different LLP states.
For these reasons, it is crucial to examine the extent to which a dedicated surface detector 
like MATHUSLA is capable of probing the parameter space of DDM scenarios. 
Indeed, the results of such a study will be presented in Sect.~\ref{sec:results}.

As we shall find, a dedicated surface detector like MATHUSLA can, in and of itself, play a 
crucial role in probing the parameter space of DDM scenarios.  
However, the prospects of correlating signal information gleaned from MATHUSLA with signal
information gleaned from the main LHC detectors renders MATHUSLA an even more powerful
tool for detecting and characterizing DDM scenarios.  For example, despite its minimal
instrumentation, analysis of observed LLP decays at MATHUSLA can not only measure the 
boost of the LLP in each individual detection event, but also strongly suggest the most 
likely decay mode for the LLP in that event on the basis of purely geometric 
information~\cite{Curtin:2017izq}.  This allows one to identify the particular HL-LHC bunch crossing 
in which the LLP was created.  On the basis of information recorded by the main detector for 
the corresponding collision event --- or the absence of such information in cases where no
L1 trigger criterion is satisfied --- the production mode of the LLP, and hence also its mass,
can then be determined or constrained.  

For DDM, the importance of these correlations 
extends beyond the characterization of any one LLP signal.  As we have discussed above, 
detailed analysis of data from MET searches at the main LHC detectors can also reveal 
information about the DDM ensemble.  Correlating information gleaned from such searches 
with information gleaned from LLP searches both at MATHUSLA and at in the main detectors 
themselves can yield further information still.
Moreover, even without a significant MET signal, the correlation of LLP searches between 
the main detectors and MATHUSLA can nevertheless reveal non-trivial information about 
the DDM ensemble.  For example, it is possible that a given DDM ensemble contains both 
shorter-lived and longer-lived LLP states.  For such an ensemble, a significant number of
displaced vertices would be expected in the main detector, while a small 
number of events --- but a number potentially sufficient to constitute 
a signal nevertheless --- would also be expected at MATHUSLA.~  Taken together, these signals 
could reveal the existence of both the shorter-lived and longer-lived portions of
the DDM ensemble.  

The enhanced prospects for probing the structure of DDM ensembles afforded by 
the correlation of signal information between the main detectors and MATHUSLA clearly merit 
further study.  In this paper, however, we shall henceforth focus on the prospects afforded by
MATHUSLA alone, and leave the analysis of how such correlations can improve the reach of LLP searches 
for future work~\cite{toappear}.

\FloatBarrier
%%%%%%%%%%%%%%%%%%%%%%%%%%%%%%%%%%%%%%%%%%%%%%%%%%%%%%%%%%%%%%%%%%%%%%%%%%%%%%%%%%%%%%
\section{The reach of MATHUSLA within the DDM parameter space:  ~A general study\label{sec:results}}
%%%%%%%%%%%%%%%%%%%%%%%%%%%%%%%%%%%%%%%%%%%%%%%%%%%%%%%%%%%%%%%%%%%%%%%%%%%%%%%%%%%%%%

In order to provide a quantitative assessment of the reach of MATHUSLA within the DDM parameter space, 
we shall conduct a toy study of the simplest situation in which our DDM dark-matter constituents $\chi_n$ are
produced through the prompt decays of a single heavy parent particle $\phi$ of mass $m_\phi$ 
via a two-body process of the form $\phi\to \overline{\chi_n} \chi_n$.    
At our level of analysis the spins of $\phi$ and $\chi_n$ are irrelevant,
and we shall imagine that 
$\phi$ itself is produced at threshold so that it is essentially at rest at the time of its decay.
We shall let $\sigma_\phi$ denote the LHC production cross section of $\phi$, and likewise we shall
assume that the $\chi_n$ masses $m_n$, 
decay widths $\Gamma_n$, and 
couplings $c_{nn}$ 
governing the $\phi \overline{\chi_n}\chi_n$ interactions
are described by the DDM scaling 
relations in Eqs.~(\ref{masses}), (\ref{decaywidths}), and (\ref{couplings}), respectively.
We therefore have a nine-dimensional parameter space
$\lbrace m_\phi, \sigma_\phi, m_0, \Delta m, \delta, \Gamma_0, y, c_0, \xi\rbrace$.

For concreteness, we take $\Gamma_0$ to be determined by the traditional dark-matter hyperstability bound, 
{\it i.e.}\/, $\Gamma_0 = (10^9 \, t_{\rm now})^{-1}$
where $t_{\rm now} = 4.35 \times 10^{17}$~s is the current age of the universe.  (Larger values will simply linearly rescale the signal in the long-lifetime limit.) We also set $m_\phi = 2$~TeV as a concrete benchmark, to be discussed further below.  
We further take $\delta = 1.5$, which lies within the range spanned by the set of theoretically-motivated benchmark values for $\delta$ discussed in Sect.~\ref{sec:DDM}.  As will become evident, shifting $\delta$ while holding the other parameters fixed will not change our fundamental results and will merely shift the portions of the DDM ensemble to which MATHUSLA is ultimately sensitive.
Likewise, the quantity $c_0$ determines the overall scale of the branching ratios ${\rm BR}(\phi\to\overline{{\chi_n}}\chi_n)$, and thus determines the total invisible branching fraction ${\rm BR}_{\chi\chi}\equiv\sum_{n=0}^\infty {\rm BR}(\phi\to\overline{{\chi_n}}\chi_n)$.  Since MATHUSLA is ultimately sensitive not to $\sigma_\phi$ alone but to the product $\sigma_\phi {\rm BR}_{\chi\chi}$, we therefore now have a seven-dimensional parameter space $\lbrace m_\phi, \sigma_\phi {\rm BR}_{\chi\chi}, m_0, \Delta m, \delta, y, \xi\rbrace$.  We shall therefore quantitatively assess the reach of the MATHUSLA detector in terms of the minimum value of $\sigma_\phi {\rm BR}_{\chi\chi}$  that gives rise to four observed LLP decays within the MATHUSLA decay volume, given specific values of the remaining six parameters $\lbrace m_\phi, m_0, \Delta m, \delta, y, \xi\rbrace$.  In the zero-background regime, this can be interpreted as an exclusion limit on $\sigma_\phi {\rm BR}_{\chi\chi}$.  In the event that LLP decays are observed, this would correspond roughly to the minimum cross section required for DDM discovery.

For any value of $m_\phi$, the decays of $\phi$ can potentially produce the ensemble 
constituents $\lbrace \chi_0,\chi_1,...\chi_{n_{\rm kin}}\rbrace$, where $n_{\rm kin}$ is the kinematic limit, defined as the maximum value of $n$ for which $m_n\leq m_\phi/2$.
Indeed, each such dark-matter constituent $\chi_n$ is 
produced with a relativistic boost factor $\gamma_n \beta_n = m_\phi \sqrt{ 1-4 m_n^2/m_\phi^2} /(2 m_n) $.
Likewise, we shall define two further quantities $n_{\rm min}$ and $n_{\rm max}$ as those values of $n$ which delimit the range of ensemble constituents $\chi_n$ whose subsequent decays into SM states are responsible for approximately 90\% of the observed events within MATHUSLA.~
Thus $(n_{\rm min}, n_{\rm max})$ describes that portion of the DDM ensemble to which the MATHUSLA detector is 
most sensitive.  
Finally, we define $n_{\rm cs}$ as indicating the heaviest ensemble constituent $\chi_n$ which is cosmologically stable, with $\tau_n\equiv \Gamma_n^{-1} \geq t_{\rm now}$.
Thus only the ensemble constituents $\lbrace \chi_0,\chi_1,...,\chi_{n_{\rm cs}}\rbrace$ will have survived to the present time and have the potential to contribute to the total present-day dark-matter abundance 
$\Omega_{\rm CDM}\approx 0.26$.
We have already noted that the MATHUSLA detector, while capable of probing large portions of the DDM ensemble, 
cannot actually probe those elements of the ensemble which constitute dark matter today.   
Thus, as a general rule, we shall always find that $n_{\rm cs} < n_{\rm min}$.   

In this connection, it is important to note that situations for which $n_{\rm cs}<1$ 
represent the ``traditional'' limit of the DDM scenario in which 
only a single dark-matter state $\chi_0$ comprises the dark matter of the universe at the present time.
Indeed, this would be the case regardless of the particular spectrum 
of abundances $\Omega_n$ that might have been generated in the early universe.
However, even in such cases, the dynamics of the universe through much of cosmological history would be
significantly different from what would be expected within a traditional single-component theory of 
dark matter, and indeed MATHUSLA --- while
not probing present-day dark matter ---  would nevertheless be probing those portions of the DDM ensemble which
are relevant for producing this non-trivial dark-sector dynamics.

%================ FIGURE ================================================
\begin{figure*}[t]
\centering
\includegraphics[width=0.32\textwidth,keepaspectratio] {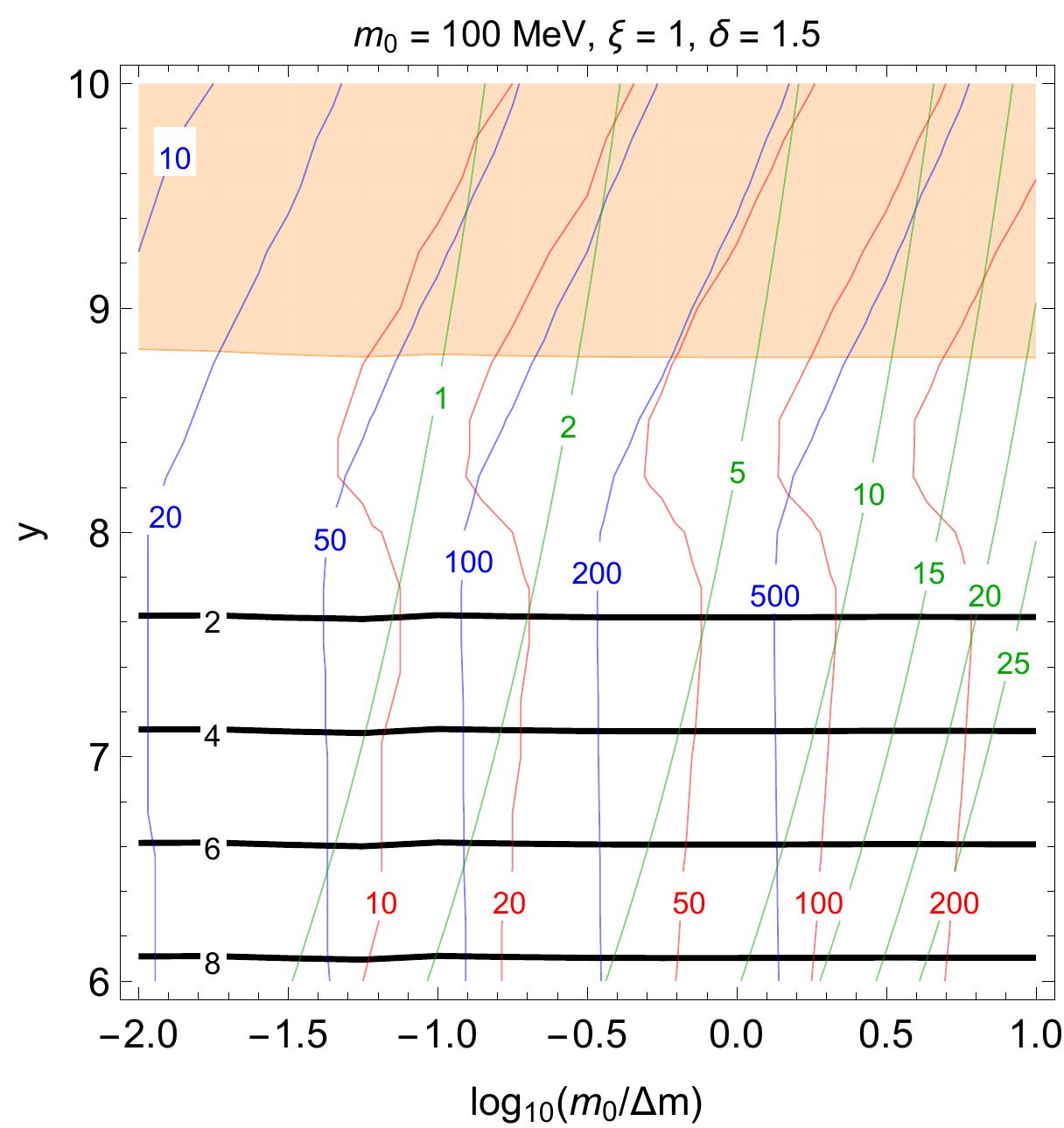}
\hfill
\includegraphics[width=0.33\textwidth,keepaspectratio] {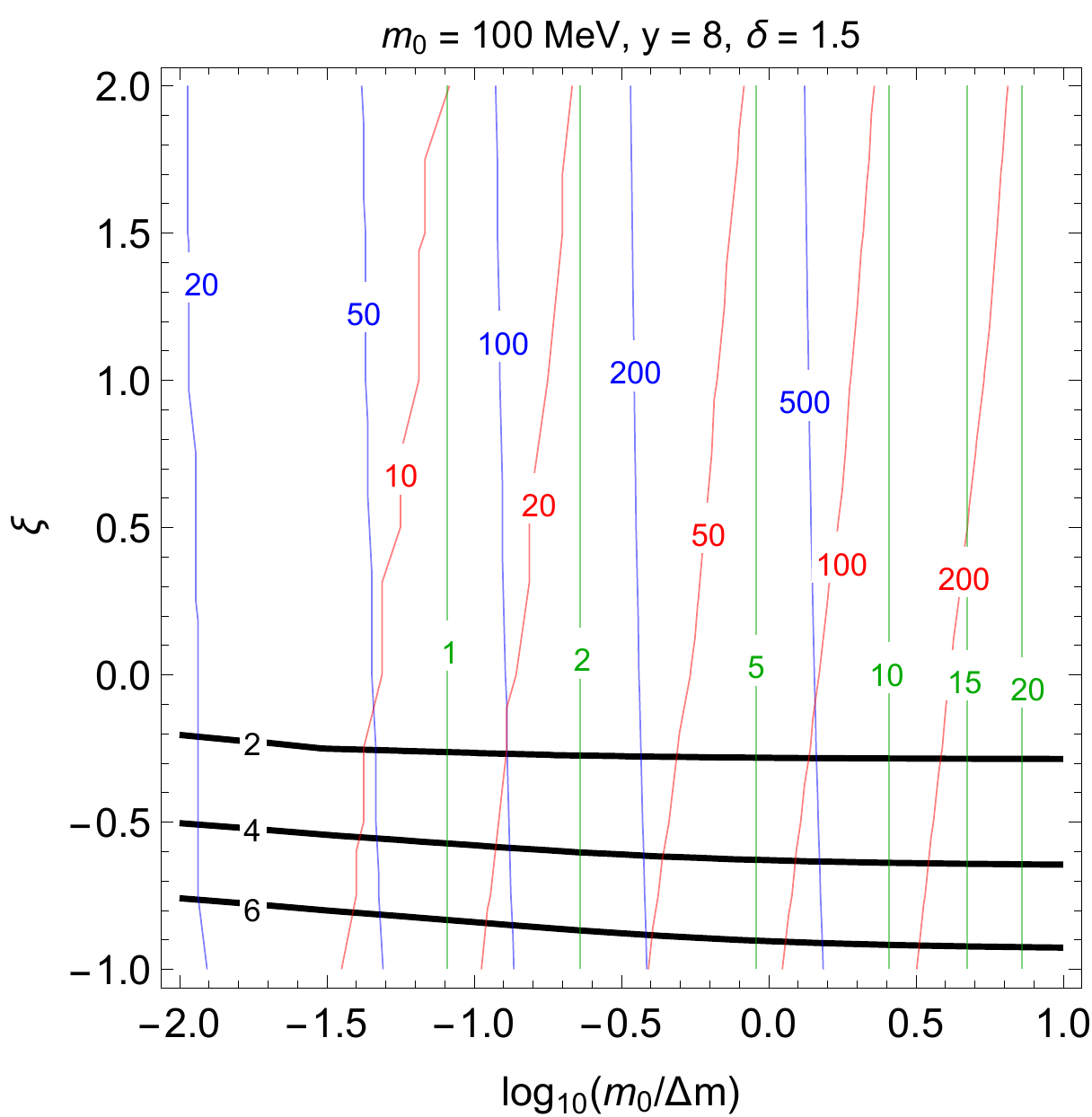}
\hfill
\includegraphics[width=0.33\textwidth,keepaspectratio] {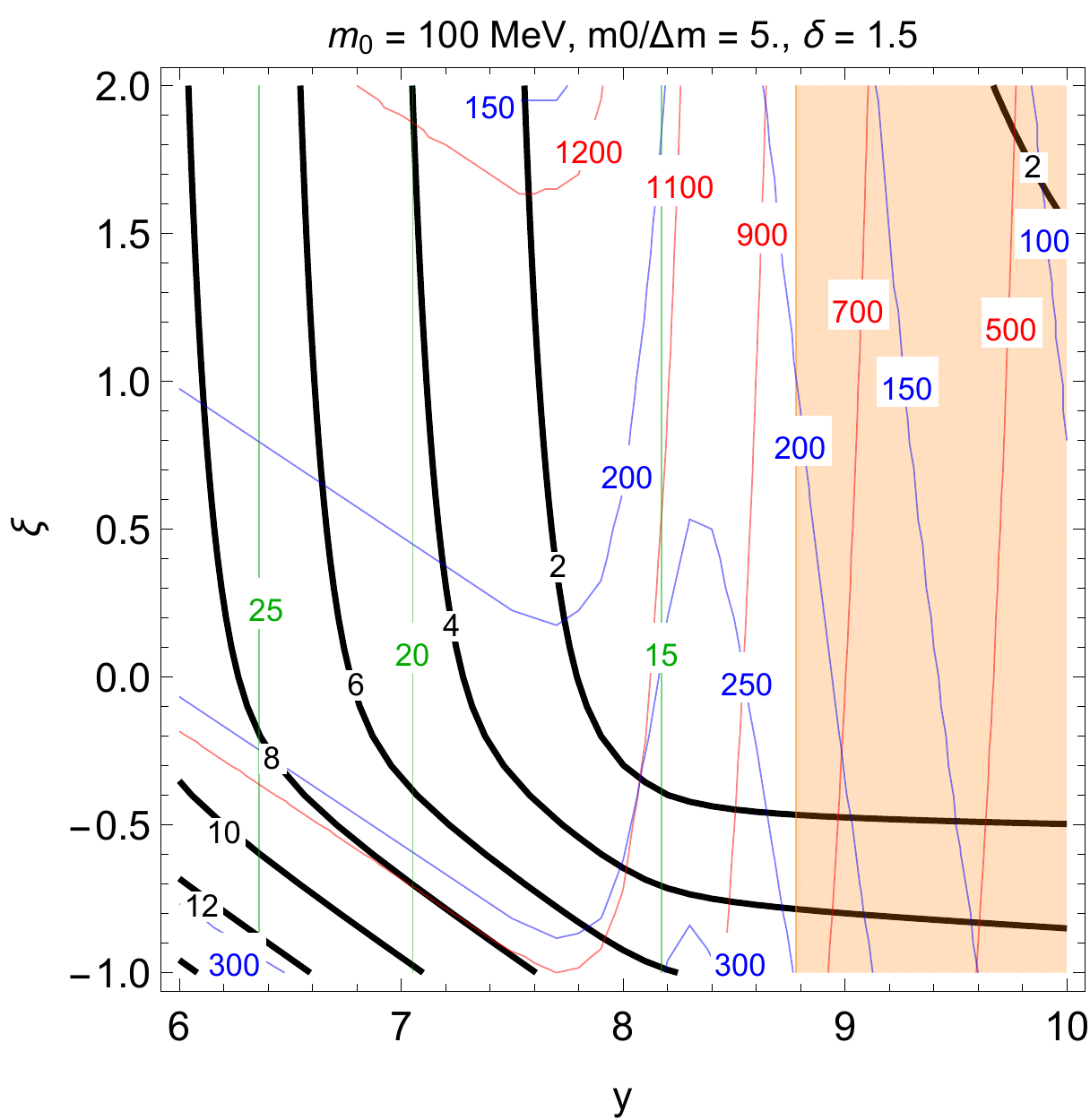}
\vskip 0.2 truein
\hskip 0.3 truein \includegraphics[width=0.40\textwidth,keepaspectratio] {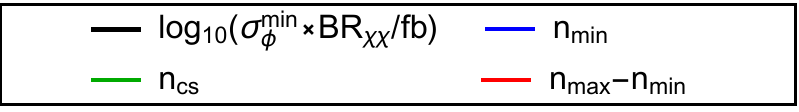}
\caption{The reach of MATHUSLA within the DDM parameter space for the benchmark values 
         $m_0=100$~MeV and $\delta=1.5$.
     Black curves indicate contours of $\sigma_\phi^{\rm min} {\rm BR}_{\chi\chi}$, while
      blue, red, and green curves indicate contours of $n_{\rm min}$, 
     $n_{\rm max}-n_{\rm min}$, and $n_{\rm cs}$, respectively.
     Likewise, the orange shading indicates the region 
        of DDM parameter space in which at least one of the ensemble constituents $\chi_n$ has a  
  characteristic decay length $\beta\gamma c\tau_{\mathrm{min}} < 1$~m.
  As discussed in the text, the region with 
     $m_0 /\Delta m \gtrsim 0.1$, $7.5 \lesssim y \lesssim 8.8$, and  $\xi \gtrsim -0.3$ 
   is a particular ``sweet spot'' within which multiple light states within the DDM ensemble 
       comprise the present-day dark matter while numerous heavier states within the same ensemble
       can lead to an observable signal at MATHUSLA.}
\label{fig:fig1}
\end{figure*}
%================ FIGURE ================================================

Given these definitions, our results are as follows.
In Fig.~\ref{fig:fig1}, we indicate the sensitivity of MATHUSLA by plotting contours (black curves)
of $\sigma_\phi^{\rm min} {\rm BR}_{\chi\chi}$, where 
$\sigma_\phi^{\rm min}$ is  the minimum production cross section for the parent particle $\phi$ 
which will produce at least four signal events within MATHUSLA.~
Within each panel we also show contours of $n_{\rm min}$ (blue curves), 
$n_{\rm max}-n_{\rm min}$ (red curves), and $n_{\rm cs}$ (green curves).  
The orange shaded regions  are the regions in which at least one of 
the $\chi_n$ has a characteristic decay length $\beta\gamma c\tau_{\mathrm{min}} < 1$~m.
In the left and center panels of Fig.~\ref{fig:fig1}, 
the contours are plotted within the $( m_0/\Delta m, y)$ and
$(m_0/\Delta m, \xi)$ planes, respectively, while in the right panel of Fig.~\ref{fig:fig1}
these contours are plotted directly within the $(y,\xi)$ plane, thereby illustrating 
the dependence on both DDM scaling exponents simultaneously.
For each plot
we have chosen the benchmark values
$m_\phi = 2$~TeV, $m_0 = 100$~MeV, and $\delta = 1.5$.
While the results in Fig.~\ref{fig:fig1} correspond to the case in which the $\chi_n$ are real scalars,
the results for spin-1/2 fermions are qualitatively similar.

In this connection, our choice of the benchmark value $m_\phi = 2$~TeV deserves further comment.  This benchmark is motivated in part by a self-consistency requirement:  in order for the ensemble to lead to a detectable signal at MATHUSLA during the HL-LHC run, the production cross section $\sigma_\phi$ must exceed the sensitivity threshold $\sigma_\phi^{\mathrm{min}}$ at any point within the DDM parameter space.  For example, if $\phi$ is a real scalar that couples to quarks through a Yukawa-type interaction with a flavor-independent coupling constant $g_q$, the dominant production process for $\phi$ is resonant production of $\phi$ through quark fusion.  In this case, we find that the product of the production cross section and this branching fraction is $\sigma_\phi \times \mathrm{BR}_{\chi\chi} \sim 100$~fb  for the choice $m_\phi = 2$~TeV (with $g_q = 0.15$ and $c_0$ chosen such that the total branching fraction $\mathrm{BR}_{\chi\chi}$ of $\phi$ to $\chi_n$ pairs is 0.5).  As $m_\phi$ increases beyond this benchmark value, $\sigma_\phi$ rapidly decreases, rendering nearly all of the DDM parameter space beyond the reach of MATHUSLA during the upcoming LHC run.  By contrast, while $\sigma_\phi$ can be significantly larger than $100$~fb for $m_\phi$ below our $2$~TeV benchmark, ATLAS and CMS searches for new physics in the monojet + MET~\cite{\ATLASMonojet,\CMSMonojet} and dijet~\cite{\ATLASDijet,\CMSDijet} channels impose stringent lower bounds on $m_\phi$.  Nevertheless, values of $m_\phi$ at or slightly below this benchmark are consistent with these constraints.  Thus, we see that the choice $m_\phi=2$~TeV corresponds to a MATHUSLA sensitivity in the range $\sigma_\phi^{\mathrm{min}} \times \mathrm{BR}_{\chi\chi} \sim 100$~fb and that values of $m_\phi$ near this benchmark are of particular phenomenological interest.

We see from the results shown in the left panel of Fig.~\ref{fig:fig1} 
that there is indeed a substantial region of parameter space within which MATHUSLA 
is capable of detecting a DDM ensemble.
Of course, the main-detector reach for our simple scenario depends strongly on the decay mode of the DDM
states, which is not specified in our toy model.   However, there are many general scenarios, such as 
decay to hadrons or Yukawa- or gauge-ordered democratic decay to SM fermions, for which MATHUSLA is likely to exceed
the main-detector reach by orders of magnitude.

%================ FIGURE ================================================
\begin{figure*}[t]
\centering
\includegraphics[width=0.32\textwidth,keepaspectratio] {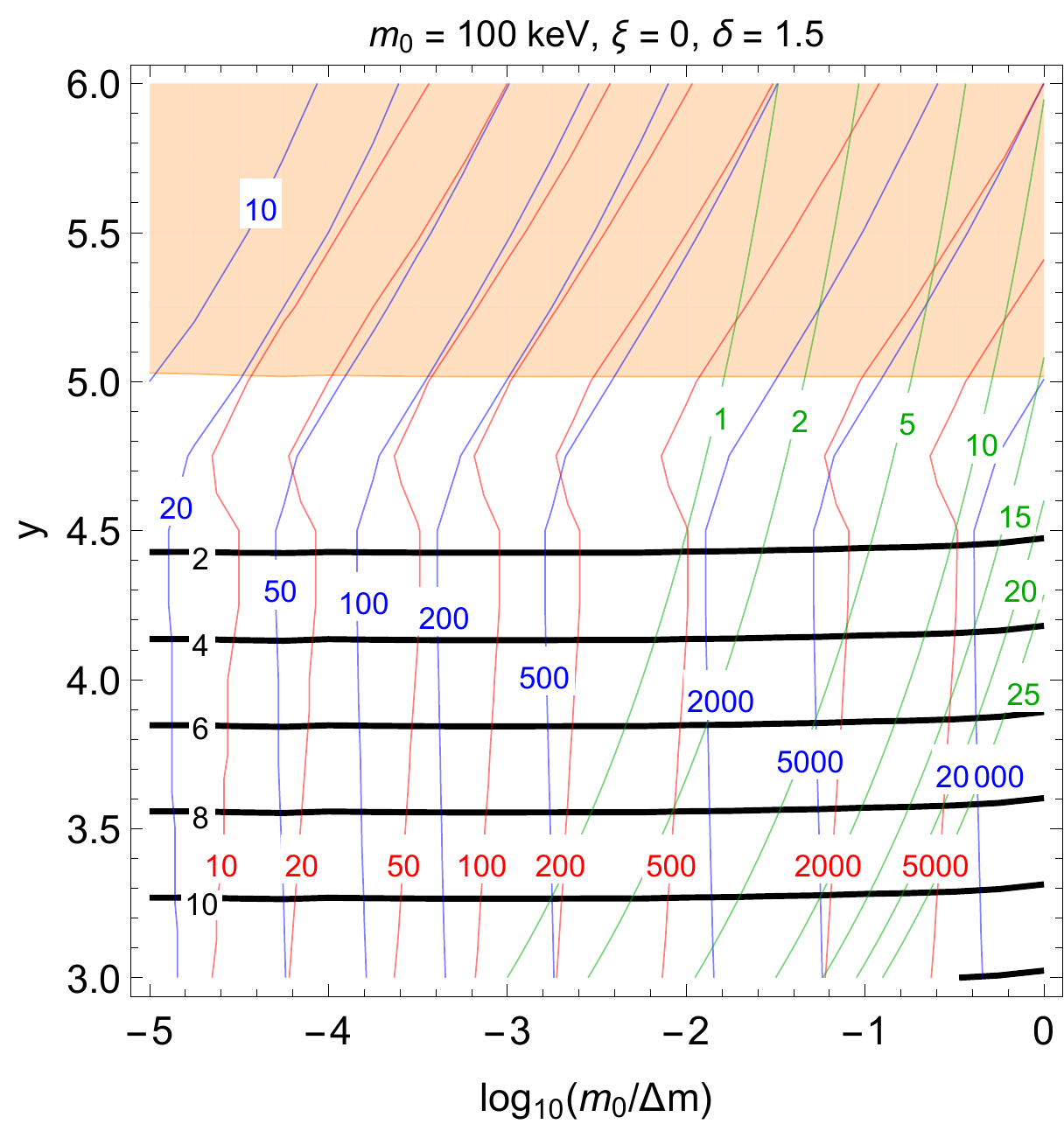}
\hfill
\includegraphics[width=0.33\textwidth,keepaspectratio] {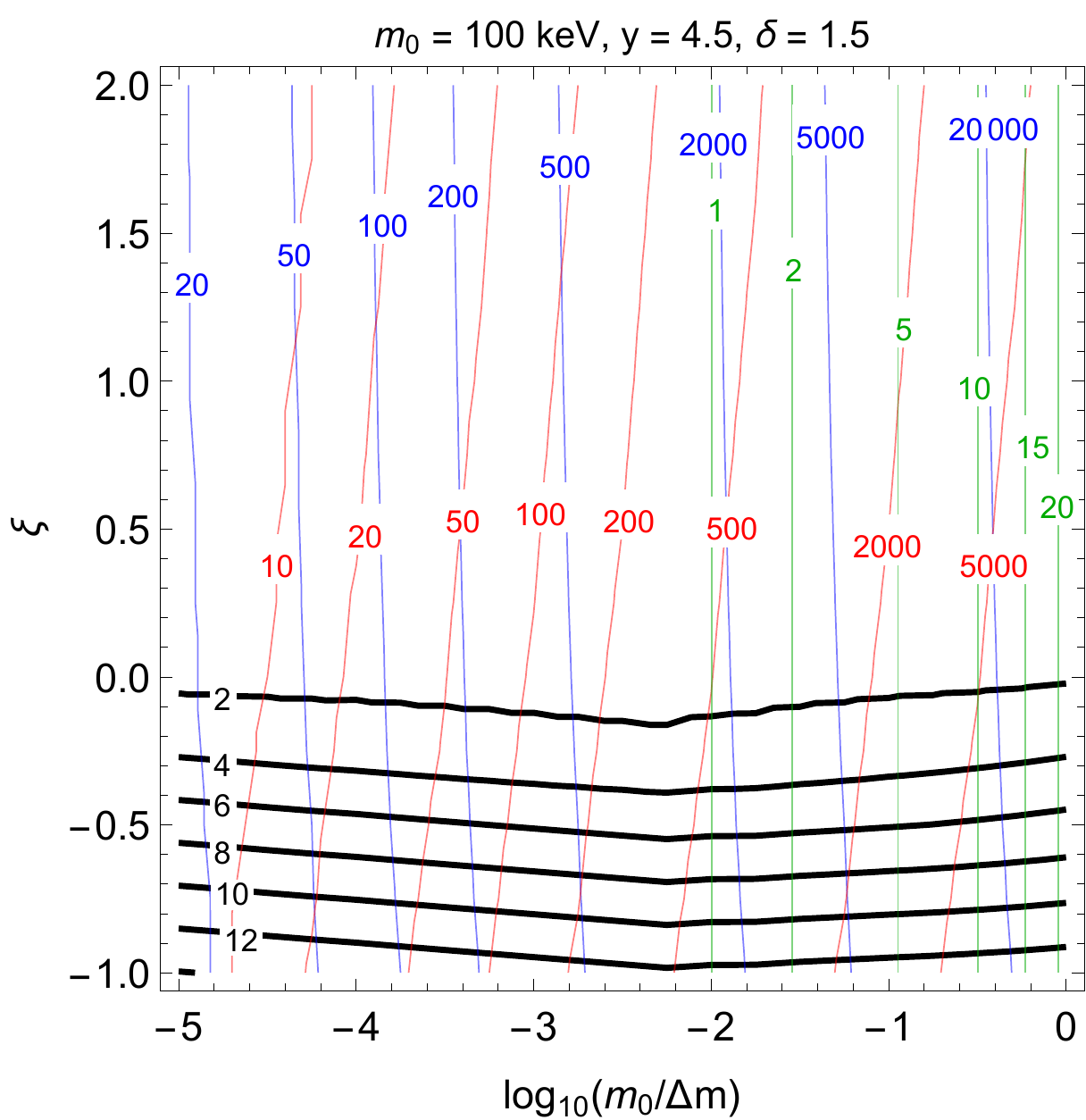}
\hfill
\includegraphics[width=0.33\textwidth,keepaspectratio] {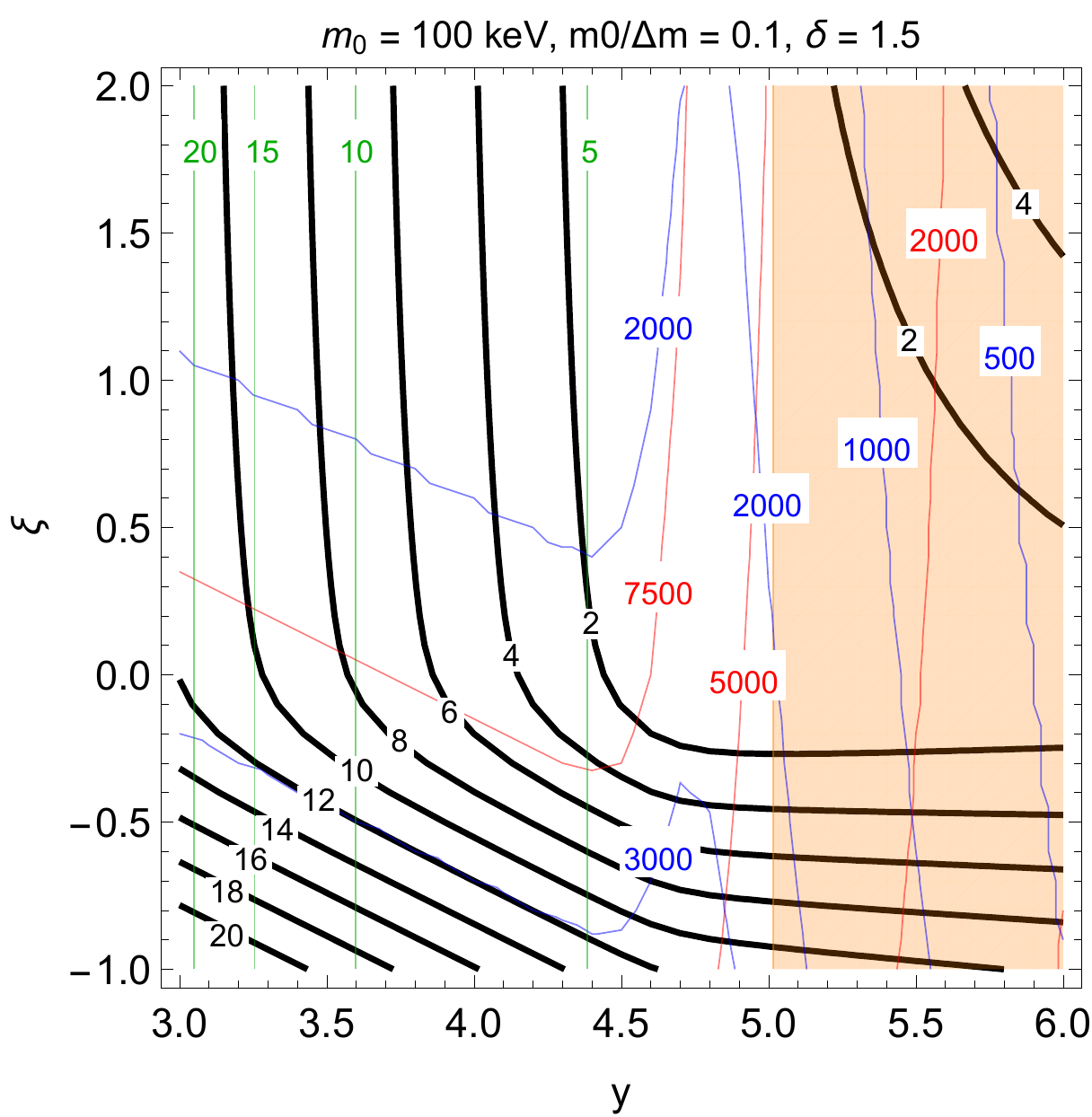}
\vskip 0.2 truein
\hskip 0.3 truein \includegraphics[width=0.40\textwidth,keepaspectratio] {plot_legend.pdf}
\caption{Same as Fig.~\ref{fig:fig1}, except that we have now shifted $m_0$ from $100$~MeV to $100$~keV.~
   This allows MATHUSLA to be sensitive to DDM ensembles with smaller values of $y$,
   leading to an even more compelling ``sweet spot'' with
      $m_0/\Delta m\gsim 0.01$,
     $4.3\lsim y \lsim 5.0$, and $\xi \gsim -0.2$.}
\label{fig:fig2}
\end{figure*}
%================ FIGURE ================================================

As indicated in the left panel of Fig.~\ref{fig:fig1}, the most compelling region of DDM parameter space
for a MATHUSLA signal 
is that within which
$7.5 \lesssim y\lesssim 8.8$.  For $y \gtrsim 8.8$, the characteristic 
decay lengths of the heaviest states in the tower fall below 
$\beta \gamma c\tau_n \lesssim \mathcal{O}(1\,\mathrm{m})$.  Since a
significant number of particles with decay lengths in this regime would decay inside 
the main collider detector, ensembles with $y \gtrsim 8.8$ would either be detected
at the HL-LHC without the help of MATHUSLA or would already have been detected 
during the current LHC run.  On the other hand, for $y \lesssim 7.5$, a parent-particle 
production cross section $\sigma_\phi {\rm BR}_{\chi\chi} \gtrsim 10^3$~fb is required in order for the expected 
number of signal events in the MATHUSLA detector to exceed the detection threshold.
This is approaching the upper range of typical strong production rates for TeV-scale states.  
Furthermore, given the sensitivity of monojet searches to invisible Higgs decays,
such large cross sections are likely to be detectable (and possibly excluded by) 
current or future 
LHC monojet searches 
or future HL-LHC monojet searches.

The center panel of Fig.~\ref{fig:fig1} indicates how the sensitivity of 
MATHUSLA depends on $\xi$, the
scaling exponent for the couplings in Eq.~(\ref{couplings}).
For this plot
we have taken a fixed scaling exponent $y = 8$ for the decay widths of the $\chi_n$.  
We see from this figure that there is generally a loss of sensitivity for MATHUSLA
as $\xi$ decreases.  This behavior ultimately reflects the fact that
for $\xi < 0$, the width of $\phi$ is dominated by decays to the lightest states in
the DDM ensemble, which are also the states with the longest lifetimes.  

We see from the left and center panels of Fig.~\ref{fig:fig1} 
that the reach of the MATHUSLA detector is 
not particularly sensitive to the ratio $m_0/ \Delta m$ --- 
at least not within the region of parameter space shown.
However, we see that this ratio nevertheless plays a crucial role
in determining $n_{\rm cs}$, the number of $\chi_n$ states
which are cosmologically stable, with $\tau_n \gtrsim \tnow$.  
Indeed, given the contours of $n_{\rm cs}$ shown in Fig.~\ref{fig:fig1},  
we see that a significant number of ensemble constituents $\chi_n$ are 
cosmologically stable for $m_0 /\Delta m \gtrsim 0.1$. 
By contrast, for $m_0/\Delta m \lesssim 0.1$, the only
contribution from the ensemble to $\OmegaDM$ is that associated with the single
lightest particle species $\chi_0$.  Thus, the region of parameter space
 in which $\xi \gtrsim -0.3$, $m_0 /\Delta m \gtrsim 0.1$, 
and $7.5 \lesssim y \lesssim 8.8$ is of particular interest from a DDM perspective,
with many individual dark-matter components $\chi_n$ potentially comprising
the total present-day dark-matter abundance $\Omega_{\rm CDM}$.

While the left and center panels of Fig.~\ref{fig:fig1} illustrate how the contours of
$\sigma_\phi^{\rm min} {\rm BR}_{\chi\chi}$, $n_{\rm min}$, 
     $n_{\rm max}-n_{\rm min}$, and $n_{\rm cs}$
depend on 
$m_0/\Delta m$ and either of our scaling exponents ($y$ for the left panel,
$\xi$ for the center panel), in the right panel of Fig.~\ref{fig:fig1} 
we have plotted these contours directly relative to both scaling exponents together. 
To do this, we have chosen the ``sweet spot'' benchmark value $m_0/\Delta m= 1$, corresponding
    to $n_{\rm cs} \approx 5$.
   We observe from this panel that the sensitivity of MATHUSLA generally depends on both scaling
exponents.  However, we also find that this sensitivity loses its dependence 
on either exponent if the other exponent becomes sufficiently large.

Taken together, the results in Fig.~\ref{fig:fig1}
indicate that there 
exists a significant region of parameter space within which multiple light states in
the DDM ensemble can contribute non-negligibly to the present-day
dark-matter abundance --- all  while heavier states in
the same ensemble can lead to an observable signal at MATHUSLA.~ 
This alone provides explicit verification that MATHUSLA can be 
particularly relevant for collider-based probes of the DDM framework.

The above results indicate the existence of a ``sweet spot'' within which the values of the scaling exponent $y$ 
are relatively large.  Although there is no fundamental reason
why such values are problematic, 
it would be interesting 
from a theoretical and aesthetic perspective
to know whether the same successes can be 
achieved with smaller values of $y$.  
Fortunately, such regions of parameter space also exist. 
In Fig.~\ref{fig:fig2}, we plot essentially the same information as we plotted in Fig.~\ref{fig:fig1},
the only significant change being that we have now taken $m_0=100$~keV rather than $m_0= 100$~MeV.~
We see that this shift in $m_0$ has not changed the gross features of these plots
relative to those in Fig.~\ref{fig:fig1},
but has shifted the regions in which MATHUSLA is most sensitive down to smaller values of $y$ --- precisely 
as desired.
Indeed, we now see that MATHUSLA remains sensitive to the DDM ensemble even below $y\approx 5$ ---
a very natural value for $y$, given that this value corresponds to a dimension-six decay operator
according to Eq.~(\ref{yd}).
Once again, just as for greater $m_0$, we find that taking $m_0/\Delta m\lsim 0.01 $ leads to situations in
which only a single dark-matter component survives to the present day.   
However, for values of $m_0/\Delta m\gsim 0.01$,
we find that  multiple components of the ensemble survive to the present day and can potentially
contribute to $\Omega_{\rm CDM}$.
Thus, for $m_0\approx 100$~keV, we see that we now have a ``sweet spot''
within the region of DDM parameter space with
      $m_0/\Delta m\gsim 0.01$,
     $4.3\lsim y \lsim 5.0$, and $\xi \gsim -0.2$.~
Of course, this region of parameter space 
corresponds to a dark sector with many light degrees of freedom.
Such a region will therefore be subject to numerous stringent astrophysical constraints.
This region is nevertheless an intriguing one, and we leave the detailed phenomenological analysis 
of this region of parameter space for future work.

\FloatBarrier
%%%%%%%%%%%%%%%%%%%%%%%%%%%%%%%%%%%%%%%%%%%%%%%%%%%%%%%%%%%%%%%%%%%%%%%%%%%%%%%%%%%%%%
\section{Discussion and Conclusions\label{sec:Conclusions}}
%%%%%%%%%%%%%%%%%%%%%%%%%%%%%%%%%%%%%%%%%%%%%%%%%%%%%%%%%%%%%%%%%%%%%%%%%%%%%%%%%%%%%%

As we have discussed, DDM is an alternative dark-matter framework in which 
the dark sector no longer consists of a single (or a few)
hyperstable dark-matter constituents, but instead consists of an entire ensemble of dark-sector states
whose lifetimes are balanced against their cosmological abundances.  
Indeed, the DDM framework arises naturally 
in a variety of top-down theoretical frameworks and
gives rise to metastable states that are related through specific scaling relations
 to the relatively stable states that constitute the dark-matter abundance today. 
Given this broad spectrum of realizable lifetimes, we have shown that MATHUSLA will be an important discovery and
diagnosis tool for DDM.~ Moreover, if all accessible ensemble states have decay lengths exceeding the main-detector
size, MATHUSLA could easily be the first or only discovery opportunity for DDM.~

Many avenues remain open for future investigation.
In this paper we analyzed the reach of the MATHUSLA detector within the DDM parameter space, 
focusing on potential LLP signals that might be observed  
within MATHUSLA alone.
However, as discussed in Sect.~\ref{sec:colliders}, there also remains the possibility of
enhancing such results by correlating LLP signal information from MATHUSLA with
both LLP and MET signal information from the main detectors --- an approach which might
be particularly fruitful in the case of DDM.~
Another direction for future study might be to analyze the implications of potential MATHUSLA data 
within the context of the complementarities that are
known to exist between different dark-matter detection methods. 
For example, within certain realizations of DDM, the interaction portal responsible for the production of the 
shorter-lived constituents in a DDM ensemble which manifest themselves at colliders as LLPs is also the portal responsible for establishing the cosmological abundances of the longer-lived, cosmologically-stable ensemble constituents which contribute to the dark-matter abundance today.  In such cases, 
information gleaned from observation of the shorter-lived constituents at MATHUSLA 
can tell us about the scaling relations which govern the spectrum of masses within the ensemble 
as well as their couplings to the mediator particles associated with that portal, and this information can then
be extended to predict the properties of the longer-lived states which constitute the dark matter today.  
In this way, data from MATHUSLA could eventually guide future efforts to explore the nature of these longer-lived 
states at other, complementary experiments which probe --- either directly or indirectly --- 
the nature of the dark matter.   
Investigations of the capabilities of MATHUSLA along both of these lines are currently underway~\cite{toappear}.

MATHUSLA can also play a complementary role in relation to other experiments either proposed or under construction which would also be capable of probing the physics of LLPs.  
Of course, MATHUSLA is ideally suited for the detection of DDM ensembles whose constituents are produced with significant transverse momentum at colliders --- for example, ensembles whose constituents are produced via the decays of other, heavier particles which serve as the primary portal between the dark and visible sectors.  
Indeed, other proposed detectors such as CODEX-b~\cite{CODEXb} could also potentially probe LLP scenarios of this sort. The long-lifetime sensitivity of these smaller detectors is much lower than that of MATHUSLA, but such detectors could probe somewhat shorter lifetimes or lower masses below the MATHUSLA reconstruction threshold.
However, for situations in which the DDM ensemble constituents are produced predominately in the forward direction --- for example, through kinetic mixing with the photon or with the neutral pion --- other instruments such as the proposed FASER detector~\cite{FASER} are likely to be more suitable for discovery~\cite{DDMFASER}.  Thus, in this way, MATHUSLA and FASER can be viewed as providing complementary coverage of DDM theory space at the LHC.

On a final note, we also observe that certain realizations of the DDM framework could also give rise to signals at experiments such as DUNE~\cite{DUNE},  SHiP~\cite{SHiP}, LDMX~\cite{LDMX}, SeaQuest~\cite{SeaQuest}, or 
Hyper-K~\cite{HyperK}.  Indeed, the interplay between the results obtained from low-energy facilities of this sort and the results obtained from high-energy facilities such as the LHC will play a crucial role in furthering 
our understanding of the structure of any hidden sectors that might yet be revealed by data.

\FloatBarrier
%%%%%%%%%%%%%%%%%%%%%%%%%%%%%%%%%%%%%%%%%%%%%%%%%%%%%%%%%%%%%%%%%%%%%%%%%%%%%%%%%%%%%%

\bigskip
\bigskip
\begin{acknowledgments}

%%%%%%%%%%%%%%%%%%%%%%%%%%%%%%%%%%%%%%%%%%%%%%%%%%%%%%%%%%%%%%%%%%%%%%%%%%%%%%%%%%%%%%

Each of the authors is proud to have spent one portion of his career 
in the United States and another portion in Canada.
The research activities of DC are supported 
by a Discovery Grant from the Natural Sciences and Engineering Research Council of Canada,
while the research activities of KRD are supported in part by the Department of Energy under Grant 
DE-FG02-13ER41976 (DE-SC0009913)
and by the National Science Foundation through its employee IR/D program.  
The research activities of BT are supported in part by the National Science Foundation under Grant PHY-1720430.
The opinions 
and conclusions expressed herein are those of the authors, and do not represent 
any funding agencies. 
 
\end{acknowledgments}

\end{document}